\documentclass[12pt]{article}
\usepackage{jcappub}
\usepackage{amsmath}
\usepackage{graphicx}
\usepackage{xcolor}

\usepackage{graphicx}
\usepackage{dcolumn}
\usepackage{bm}
\usepackage{hyperref}
\usepackage[utf8]{inputenc}
\usepackage[capitalise]{cleveref}
\usepackage{color}
\usepackage{aas_macros}

\newcommand{\ldb}{\lambda_\mathrm{dB}}
\newcommand{\eqdot}{\,\,.}
\newcommand{\eqcom}{\,\,,}

\abstract{In a broad class of scenarios, inflation is followed by an extended era of matter-dominated expansion during which the inflaton condensate is nonrelativistic on subhorizon scales. During this phase density perturbations grow to the point of nonlinearity and collapse into bound structures.  This epoch strongly resembles structure formation with ultra-light axion-like particles. This parallel permits us to adapt results from studies of cosmological structure formation to describe the nonlinear dynamics of this post-inflationary epoch. We show that the inflaton condensate fragments into ``inflaton clusters'', analogues of axion dark matter halos in present-day cosmology. Moreover, solitonic objects or ``inflaton stars'' can form inside these clusters, leading to density contrasts as large as $10^6$ in the post-inflationary universe.} 

\author{Jens C. Niemeyer${}^{1,2}$}
\author{Richard Easther${}^{2}$}

\emailAdd{jens.niemeyer@phys.uni-goettingen.de}
\emailAdd{r.easther@auckland.ac.nz}

\begin{document}

\title{Inflaton Clusters and Inflaton Stars}

\affiliation{%
 $^1$Institut f\"ur Astrophysik, Universit\"at G\"ottingen, Germany\\
 $^2$Department of Physics, University of Auckland, Private Bag 92019, Auckland, \\New Zealand
}%

\date{\today}

\maketitle
\section{Introduction}

Cosmological structure formation begins in earnest with the onset of matter domination, tens of thousands of years after the Big Bang. However, the Universe can pass through a much earlier phase in which initially small overdensities grow gravitationally, leading to the formation of collapsed structures on spatial scales far smaller than the prevailing Hubble horizon \cite{Easther2011DelayedOscillations,Jedamzik:2010dq,Musoke:2019ima}. This occurs during the so-called primordial dark age  \cite{Boyle:2005se}  following inflation in models which do not undergo resonance  or prompt thermalization \cite{Kofman:1994rk,Shtanov:1994ce,Kofman:1997yn,Lozanov:2016hid}, and the coherently oscillating inflationary condensate is fragmented by growing inhomogeneities. 

It was recently demonstrated that this fragmentation is well-described by the non-relativistic Schrödinger-Poisson  equation  \cite{Musoke:2019ima}, which also governs the evolution of cosmological structure in fuzzy (or axion or ultralight) dark matter models \cite{Hu:2000ke, Schive:2014dra,Marsh:2015xka, Hui:2016ltb}. From the perspective of the Schrödinger-Poisson equation, the two epochs differ only in their  initial spectra and parameter values. Consequently, it is possible for the rich phenomenology of axion clusters and Bose stars to be replicated in the early universe.  Moreover, since fuzzy dark matter behaves like generic cold dark matter on scales much larger than its de Broglie wavelength,  analytical tools that describe structure formation such as the Press-Schechter formalism \cite{Press1974}  can be adapted to the quantify the growth of these transient, primordial structures.  

We use these parallels to analyse the formation of nonlinear structures following quadratic inflation. We infer that  the Universe can become dominated by a hierarchy of gravitationally bound ``inflaton clusters'' during the primordial dark age. Further, if the matter dominated phase lasts long enough,  ``ınflaton stars'', a class of Bose stars, can form within these clusters via Bose-Einstein condensation following condensate fragmentation  \cite{Tkachev1986,Levkov2018GravitationalRegime}. 

We estimate the formation time, size and density of inflaton clusters and use the Press-Schechter formalism to obtain their halo mass function. This is in contrast to Ref~\cite{Musoke:2019ima}, which directly simulates the breakdown of the condensate  with a non-canonical initial spectrum. We show that the formation of inflaton clusters leads to overdensities roughly 200 times larger than the prevailing average density, while axion stars can lead to overdensities on the order of $10^6$. 

\smallskip

\begin{figure}
\centerline{ \includegraphics[width=.8 \textwidth]{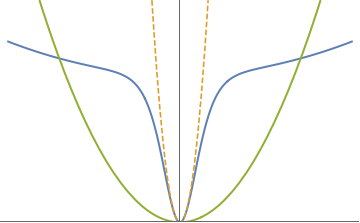}}
\caption{Two inflationary potentials: a pure quadratic potential, and an asymptotically quadratic potential, with the effective quadratic minimum shown as a dashed line.} 
\label{fig:plotsketch}
\end{figure}

\section{Setup} 
We are interested in the post-inflationary evolution of scenarios which do not undergo resonance or prompt reheating. We assume that the inflaton evolves in a quadratic minimum, or
\begin{equation}
    V(\phi) = \frac{m^2}{2} \phi^2 \,.  
\end{equation}
\Cref{fig:plotsketch} shows two potentials; the pure quadratic case and an asympotically quadratic potential. In  more complicated potentials the amplitude of small scale perturbations can differ substantially from those at present-day astrophysical scales but in what follows we focus on the pure quadratic case as an illustrative example.  This model has just one free parameter, so setting $m$ to match the inferred amplitude of the primordial spectrum at astrophysical scales  fixes the amplitude of the perturbations that  cross the horizon as inflation ends. 

Quadratic inflation ends when $\phi \sim M_p$, where $M_p$ is the reduced Planck mass. The first Friedmann equation specifies $H^2 = 8\pi G \rho/3$, where $\rho$ is the density, $G = 1/8\pi M_p^2$, so inflation ends with $H\sim  m$. Pure quadratic inflation produces a gravitational wave background that is inconsistent with current constraints, so the potential must ``flatten out'' at $\phi> M_p$. However,  most large field models can be locally approximated as $V(\phi)\sim \phi^n$, where $n$ is a parameter that is close to unity and which varies weakly with $\phi$.  Slow roll inflation ends when the parameter $\epsilon \approx 1$, or
\begin{equation} 
\frac{M_p^2}{2} \left(\frac{V'}{V} \right)^2  \approx 1 \, .
\end{equation}
Consequently, the endpoint of inflation is generically at $\phi\sim M_p$  with $H$ of order $m$ in large field scenarios, so the post-inflationary dynamics of quadratic inflation will be representative of a much larger class of models\footnote{Small field scenarios where inflation ends at a near-discontinuity in the potential will need to be treated separately; in these cases the de Broglie wavelength and Hubble radius at the end of inflation can differ significantly.}.

\section{Linear perturbation growth}
Small perturbations on subhorizon scales are governed by Newtonian perturbation theory. 
Density perturbation amplitudes at large wavenumbers $k \gg k_f$, where $k_f = a_f H_f$ is the horizon scale at the end of inflation, are strongly suppressed. This can be understood in terms of the so-called ``quantum Jeans scale'' that appears in the linear analysis of nonrelativistic scalar fields.
Describing the inflaton field in terms of a fluid with a wavenumber-dependent effective sound speed, the problem is equivalent to the Jeans analysis of gravitationally unstable fluids \cite{Hwang2009}. It follows that modes above the scalar field Jeans wavenumber $k_J$ are gravitationally stable and oscillatory, where $k_J$ is given by \cite{Hu:2000ke} 
\begin{align}
    \label{eq:kJeans}
    k_J =a (16 \pi G  \rho)^{1/4} \,\left(\frac{m}{\hbar}\right)^{1/2} \, .
\end{align}
Here, $a$ is the scale factor and we show $\hbar$ explicitly to simplify numerical calculations. At the end of inflation, $k_J$ is similar to the de Broglie wavelength and of the same order as the horizon scale $k_f$. During linear evolution, the growth of density perturbations is suppressed for modes with $k>k_J$. Conversely, longer modes with $k<k_J$ behave like standard non-relativistic matter with amplitudes proportional to the scale factor. 

A density perturbation $\delta_k$ begins to grow when $k=k_J$, i.e. at the scale factor $a(k) = a_J \sim k_J^4$. From 
\[
\frac{a}{a_J} = \frac{a}{a_f} \left(\frac{k_J}{k_J(a_f)}\right)^{-4}
\]
follows that modes with $k \gtrsim k_J(a_f)$ are suppressed with respect to standard CDM by a factor of $(k_J/k_J(a_f))^{-4}$ and unchanged for $k\lesssim k_J(a_f)$. We therefore expect density perturbations to scale as $\delta_k^2 \sim k^{-8}$, and the density power spectrum $\Delta_m^2 \sim k^3 \delta_k^2 \sim  k^{-5}$, for $k \gg k_f$.

\begin{figure}
\centerline{\includegraphics[width=.8 \textwidth]{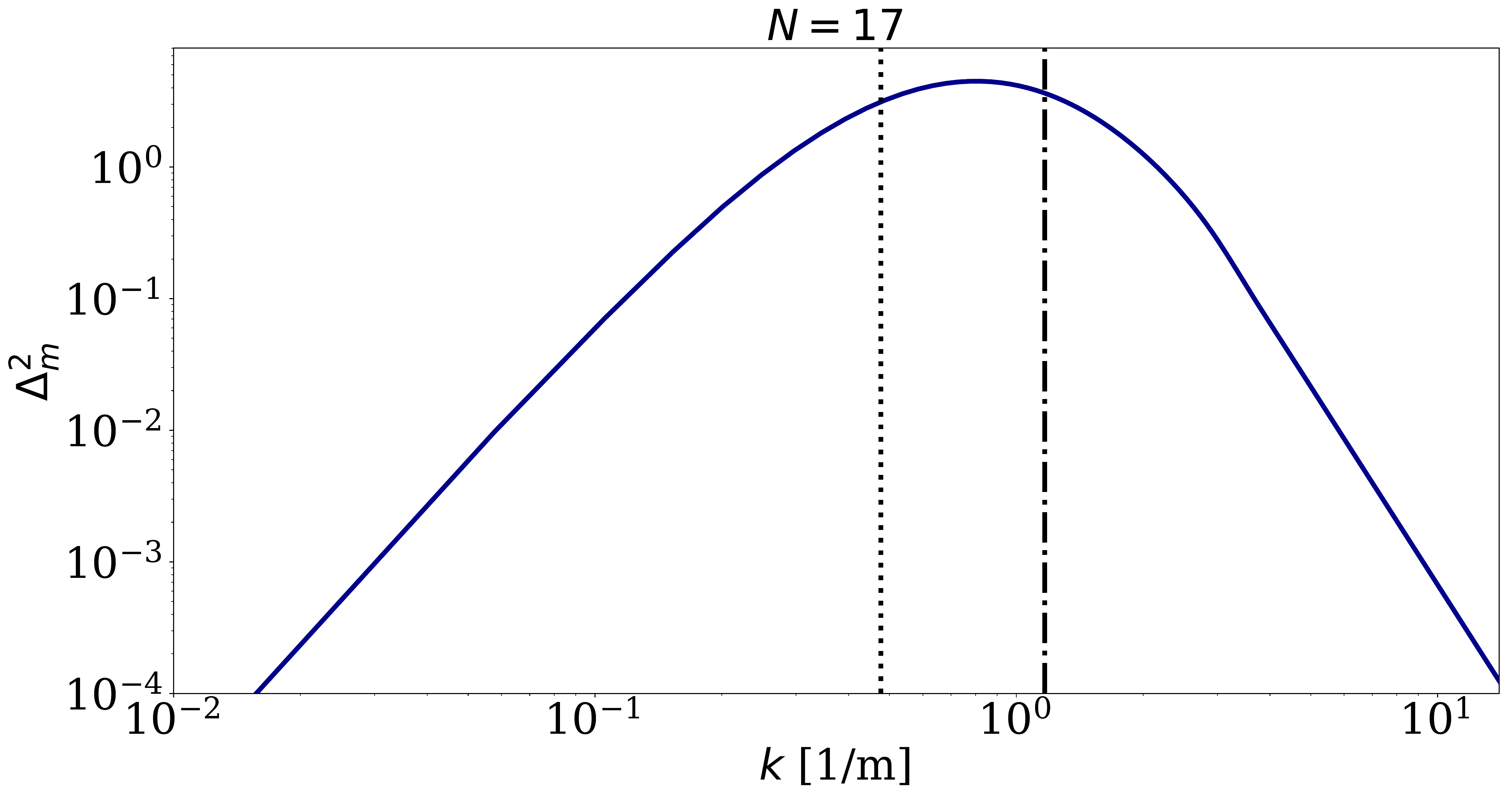}}
\caption{Linear density power spectrum $17$ $e$-folds after the end of inflation for $m^2\phi^2$-inflation with $m = 6.35 \times 10^{-6}\,M_p$ and $k_\ast = 2 \times 10^{-3}$ Mpc$^{-1}$ as in \cite{Easther2011DelayedOscillations}. The dotted and dashed-dotted lines mark the horizon and Jeans scales at the end of inflation, respectively.} 
\label{fig:plot1}
\end{figure}

The detailed linear power spectrum of the density perturbations  inside the horizon during the oscillatory phase following quadratic inflation was derived in Ref.~\cite{Easther2011DelayedOscillations} without reference to a Jeans-type analysis.  Their result maps to the standard, weakly $k$-dependent slow-roll result on scales outside the horizon at the end of inflation and drops like the expected power-law on subhorizon scales, 
$\Delta_m^2 \sim  k^{-5}$.  Since the nonlinear subhorizon evolution of perturbations does not depend on the details of the initial  spectrum \cite{Musoke:2019ima} we work with a generic spectrum, sketched in  \cref{fig:plot1}, rather than anchoring our analysis to a specific model. 

Using the same parameters as \cite{Easther2011DelayedOscillations} ($m = 6.35 \times 10^{-6}\,M_p$, $k_\ast = 2 \times 10^{-3}$ Mpc$^{-1}$) to facilitate comparison, the amplitude of horizon-scale density perturbations is of order $\Delta_m \sim 10^{-7}$.  Given that perturbations grow linearly with the scale factor, nonlinearity will set in when when the Universe has become roughly $a_c/a_f \sim 10^7$  times larger since the end of inflation, or by around 17 $e$-folds. \Cref{fig:plot1} shows the linear matter power spectrum $\Delta_m^2$ evolved to $17$ $e$-folds after the end of inflation by $(a_c/a_f)^2$.
The density perturbations peak at the comoving horizon scale at the end of inflation, $k_f^{-1}$. 

\section{Inflaton clusters} 
On scales much greater than the de Broglie wavelength $\ldb \sim h (mv)^{-1}$ the nonlinear evolution is indistinguishable from collisionless cold dark matter as a consequence of the well-known parallel between the kinetic description of random waves obeying the Schr\"{o}dinger-Poisson equations and the Vlasov-Poisson equations for collisionless systems \cite{Widrow:1993qq,Uhlemann2014SchrodingerDust}. 

\begin{figure}
\centerline{\includegraphics[width=.8 \textwidth]{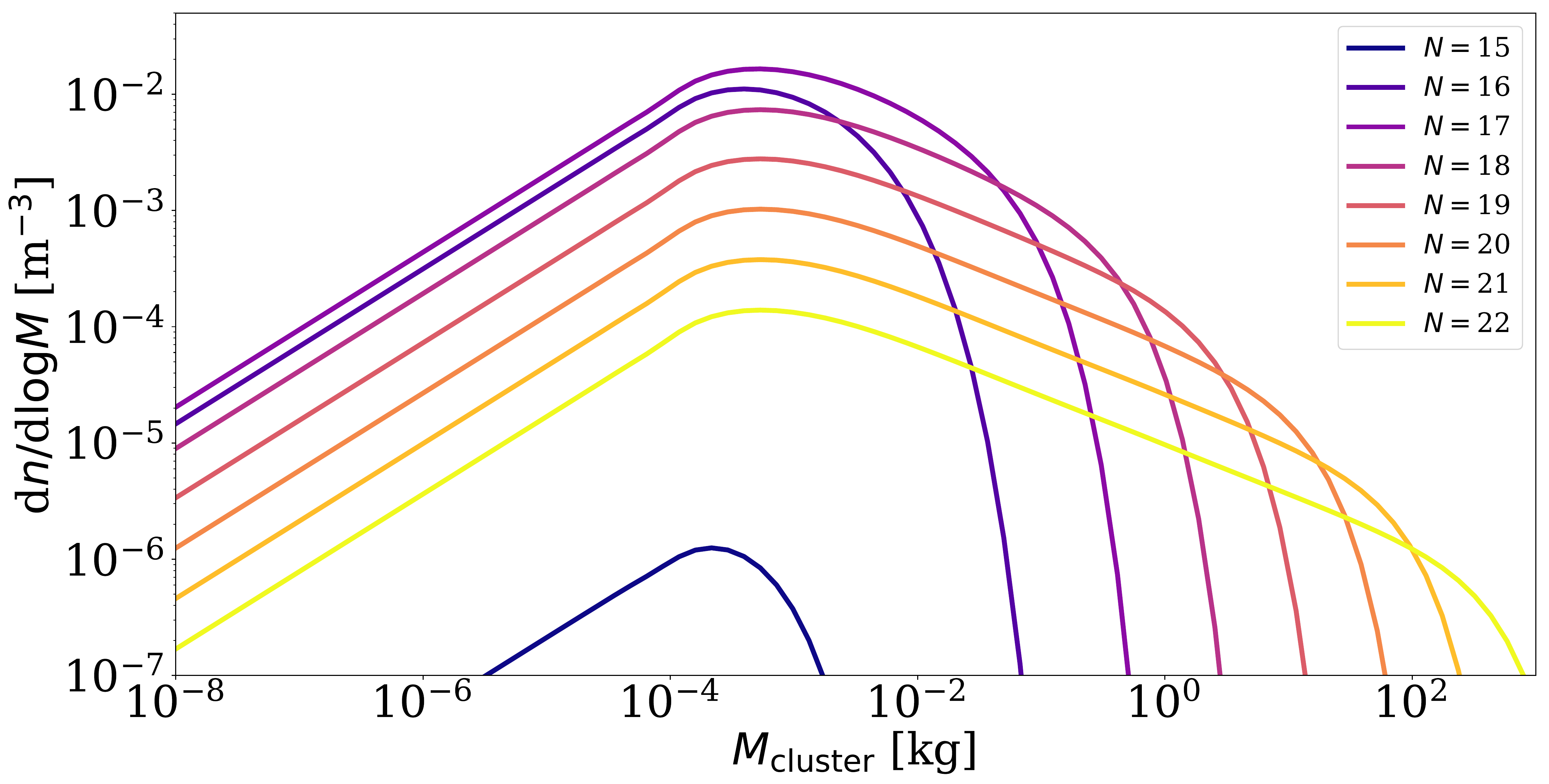}}
\caption{Press-Schechter mass functions for the Jeans-suppressed power spectrum using a sharp-$k$ filter \cite{Schneider2015StructurePerturbations} for the mass variance for different $e$-foldings $N$ after the end of inflation.} 
\label{fig:plot2}
\end{figure}
After becoming nonlinear, overdensities (which by this time are well inside the horizon) collapse on time scales of order of the dynamical time $t_d \sim (4 \pi G \rho)^{-1/2}$ {$\sim 10^{-25}\,$s.} Since $\ldb \ll a_c/k_f$ at this time (see below), we can safely neglect wave-like dynamics and apply the standard methods for collisionless collapse. Accordingly, collapsed structures virialize with typical overdensities of $\sim 18\pi^2$ into halos that we will refer to as ``inflaton clusters'' \cite{Peebles1980}. Their characteristic mass is determined by the mass inside of a comoving volume with radius $k_f^{-1}$,
\begin{equation}
    \label{eq:Mcluster}
    M_f \simeq \left(\frac{4 \pi}{3}\right)\, \left(\frac{a_f}{k_f}\right)^3\, \rho(a_f) \sim 10^{-2}\,\mathrm{kg} \eqcom
\end{equation}
and their virial radius and velocity are given by
\begin{align}
    R_\mathrm{vir} &\simeq  \left(\frac{4 \pi}{3}\right)^{-1/3}\, \left(\frac{M_f}{18 \pi^2\,\rho(a_c)}\right)^{1/3} \sim 10^{-22}\,\mathrm{m} \cr 
    v_\mathrm{vir} &\simeq \left(\frac{G M_f}{R_\mathrm{vir}}\right)^{1/2} \sim  10^4 \,\mathrm{m}\,\mathrm{s}^{-1}
\end{align}
for our choice of parameters. Computing the de Broglie wavelength as $\ldb = h/m v_\mathrm{vir}$, we verify that $\ldb k_f/ a_c \sim 10^{-4}$. However, the Schwarzschild radius from \cref{eq:Mcluster} is $\sim 10^{-7}\,R_\mathrm{vir}$, so these nonlinearities are very unlikely to lead directly to black hole formation.

The halo mass function describes the comoving number density of clusters per logarithmic mass interval $dn/d\log(M)$ and can be estimated with the Press-Schechter method \cite{Press1974}. We account for the Jeans-scale suppression of low-mass clusters by computing the mass variance using the sharp-$k$ cutoff \cite{Schneider2015StructurePerturbations} applied to the Jeans-suppressed power spectrum. Clustering sets in when perturbations have crossed the critical amplitude for collapse ($N \sim 17$ $e$-foldings after the end of inflation in our example) and proceeds by a hierarchical production of higher-mass clusters out of smaller ones, see \cref{fig:plot2}. 

As expected, the resulting initial distribution of inflaton cluster masses is strongly peaked near the horizon mass at the end of inflation, $M_\mathrm{ic} \sim 0.1\,M_f$. The initial generation of clusters will undergo mergers, leading to the production of ``halos'' that are far more massive than the initial population.  The detailed form of the halo mass function will depend on the specifics of the inflationary model. However, as noted above the nonlinear evolution only depends weakly on the initial sub-horizon spectrum, so the suppression at low masses seen in \cref{fig:plot2} is generic for  scenarios in which the post-inflationary condensate undergoes gravitational collapse. 

\section{Inflaton stars} 
On scales $\sim \ldb$ we see additional phenomena, relative to our expectations from the analogy with simple cold dark matter. In particular, the Schrödinger-Poisson equation \cite{RUFFINI1969SystemsState,Seidel1994} predicts the formation of gravitationally bound solitonic objects, or Bose stars. These are described by stationary solutions of the Schrödinger-Poisson equation and are supported against gravitational collapse by gradient energy of the scalar field.

In the context of axion dark matter it is  well established  that Bose stars can form either spontaneously by gravitational Bose-Einstein condensation \cite{Tkachev1986,Tkachev1991} or during the violent relaxation phase of collapsing bosonic dark matter halos \cite{Schive2014, Veltmaat2018FormationHalos, Eggemeier2019PhR}. The former process is entirely classical and can be described by a kinetic formalism analogous to the Landau equation \cite{Levkov2018GravitationalRegime}. Numerical simulations indicate that both the formation and mass growth of Bose stars are governed by the gravitational condensation time, which for a virialized cluster with crossing time $t_\mathrm{cr} \sim R_\mathrm{vir}/v_\mathrm{vir} \sim t_d$ can be written as $\tau \sim 0.05 \, t_\mathrm{cr}\, (\ldb/R_\mathrm{vir})^3$ \cite{Levkov2018GravitationalRegime} where we have neglected the Coulomb logarithm. 

For our parameters, $\tau \sim 10^{-19}$ s, far greater than the Hubble time at the time of collapse, $H^{-1}(a_c) \sim 10^{-26}$ s. Consequently, the formation of ``inflaton stars'' by Bose-Einstein condensation is not guaranteed and will depend on the time at which reheating occurs. However,  simulations of axion star formation in axion miniclusters \cite{Eggemeier2019PhR} suggest that their formation time might be significantly shortened by strong fluctuations of the gravitational potential during the collapse of the inflaton cluster.

After an inflaton star has formed, its mass grows in time as $(t/\tau)^{1/2}$ \cite{Levkov2018GravitationalRegime,Eggemeier2019PhR}. As first shown in simulations of ultra-light axion halos by Schive et al. \cite{Schive2014b} (and confirmed by \cite{Veltmaat2018FormationHalos}), the final mass of solitonic objects in virialized clusters follows a scaling relation with the cluster mass as $M_\mathrm{is}\sim M_\mathrm{ic}^{1/3}$. Heuristically, it follows from demanding that the soliton radius is of the order of the de Broglie wavelength of the cluster, $R_\mathrm{is} \sim \ldb$. It was shown to be a consequence of the saturation of mass growth when the virial velocity of the soliton itself begins to dominate the condensation time \cite{Eggemeier2019PhR}. Applying this relation to inflaton stars forming in inflaton clusters, we obtain typical masses of the order of 
\begin{align}
    M_\mathrm{is} &= \left(\frac{\hbar}{m}\right)\,\left(\frac{3 }{10\,a\,G}\right)^{1/2}\, \left(24 \pi^3\,\rho(a_c)\right)^{1/6}\,M_\mathrm{ic}^{1/3} \cr 
                &\sim 10^{-6}\, \mathrm{kg} \eqdot
\end{align}
Their central density is approximately
\begin{align}
    \label{eq:rho_c}
    \rho_c &\simeq 4 \times 10^{-3} \, \left(\frac{Gm^2}{\hbar^2}\right)^3\,M^4 \cr
            &\sim  10^{67}\, \mathrm{kg} \,\mathrm{m}^{-3} \sim 10^6 \, \rho(a_c)\eqdot
\end{align}

Once the Universe is fully thermalised, the density and temperature are related by 
\begin{equation}
    \rho = g_\star \frac{\pi^2}{30} T^4
\end{equation}
where $g_\star$ is the effective number of relativistic degrees of freedom. If reheating is instantaneous, the maximal temperature possible is around that will be attained is around $T_{\rm max} \approx 10^{16}\,g_\star^{-1/4}$ GeV. Standard Model particles alone contribute around 100 degrees of freedom, and at  GUT scales $g_\star$ could be much larger, but its impact is  blunted by the fractional exponent. 

Recalling that $\rho \sim a^{-3}$ during matter domination,  if  $g_\star$ is fixed we have
\begin{equation}
\frac{T_{\rm r}}{T_{\rm max}}  = \left( \frac{a_f}{a_{\rm  r}}\right)^{3/4} \, .
\end{equation}
Conseqently, if inflaton clusters form, the maximal possible reheat temperature is around $10^{-6}\, T_{\rm max}$. If the universe grows by a further factor of $(10^{-17}/10^{-25})^{2/3}$ before inflaton stars form, the maximal reheat temperature is $10^6\,g_\star^{-1/4}$ GeV, still far above LHC scale physics. 

\section{Conclusions and Discussion} 
Simulations confirm that gravitationally bound structures form after the nonlinear collapse of the inflaton condensate \cite{Musoke:2019ima}. This process obeys the same dynamics as halo formation in dark matter scenarios based on ultralight bosonic particles. Using semi-analytical tools and  rescaling simulations of axion dynamics we predict that without prompt reheating, large field inflation can be followed by the formation of a hierarchy of virialized halos (inflaton clusters) and solitons (inflaton stars), via Bose-Einstein condensation. 

If the reheating temperature is below $\sim 10^5$~GeV we  expect overdensities on the order of $10^6$. These estimates will need to confirmed with detailed simulations but it is clear that, in the absence of mechanisms for insuring prompt reheating, thermalisation will occur in a highly inhomogeneous universe, dominated by nonlinear, gravitationally bound remnants of the inflaton field. The lifetimes of inflaton clusters and stars depend on their couplings to other particles. Interestingly, inflaton stars locally restore the coherence of the inflaton field and can therefore decay by parametric resonance, analogously to  the resonant decay of axion stars into photons \cite{Tkachev1986,Hertzberg2018DarkPhotons,Sigl2019}. 

The details of our analysis do not depend directly on the inflationary potential; the phase we consider occurs long after inflation ends and the amplitude of the underlying field is much smaller than its inflationary value. In this regime, the potential is likely to be well-approximated by the first term of the Taylor series about its minimum and  can thus be treated as purely quadratic. However, to satisfy constraints on the tensor:scalar ratio the inflationary segment of the potential must be sub-quadratic in simple slow-roll scenarios. If the non-quadratic terms are significant in the early post-inflationary phase these can induce resonance (and oscillon formation) \cite{Lozanov:2016hid},  disrupting the condensate before any gravitational collapse could occur. Secondly, if inflation ends with a very sharp transition in the potential the mass of inflaton field at the origin can be significantly larger than the value derived from purely quadratic inflation. In this case the post-inflationary horizon size need not be similar to the Jeans length (as it is for the pure quadratic case) and in extreme cases this could modify the final parts of the inflationary power spectrum, which would change the details of our analysis. In extreme cases, the spectrum can increase rapidly if the inflaton approaches a near plateau, possibly leading to primordial black hole formation, which would also disrupt the condensate in ways not anticipated by our analysis. Finally, we have not considered scenarios such as Higgs-inflation \cite{Bezrukov:2007ep} in which the inflaton has non-trivial couplings to gravity; these clearly have the capacity to change the post-inflationary evolution of inflaton perturbations.

Beyond the intrinsic interest of a  nonlinear phase in the early Universe, the detailed dynamics of thermalisation determines the ``matching'' between present-day and primordial scales, which depends on the reheating history and has a small but potentially detectable impact on the observable perturbation spectrum~\cite{Dodelson:2003vq,Liddle:2003as,Peiris:2008be,Adshead:2010mc}. Moreover, the dark matter component of the present-day universe must originate in the very early Universe, and almost certainly involves physics outside the Standard Model. 

The production of dark matter and the thermalisation history of the Universe could conceivably be linked \cite{Chung:1998ua,Liddle:2006qz,Easther:2013nga,Fan:2013faa,Tenkanen:2016jic,Tenkanen:2016twd,Hooper:2018buz,Almeida:2018oid,Tenkanen:2019cik}. At the  very least, it is certain that in the absence of prompt thermalisation, reheating must take place within the highly fragmented and inhomogeneous condensate and this process is currently almost entirely unexplored. Another  fruitful area for further investigation is the possible formation of small primordial black holes \cite{Anantua:2008am,Zagorac:2019ekv,Martin:2019nuw,Muia:2019coe} as a result of inflaton interactions and the production of gravitational waves \cite{Jedamzik:2010hq} as result of inflaton star mergers and the accelerated motion of inflaton remnants in the nonlinear universe.   

Finally, even the best large scale structure simulations  struggle to follow the Universe through between one and two orders of magnitude of growth. Conversely, the Universe could conceivably grow by 10 or more orders of magnitude during the primordial dark age. Consequently, it is likely that semi-analytic estimates -- many of which can be adapted from analyses of large scale structure and boson star formation in the present epoch -- will be critical to understanding the dynamics of long matter-dominated phase in the aftermath of inflation.

\section*{Acknowledgements}
We thank Benedikt Eggemeier, Mateja Gosenca, Lillian Guo, Shaun Hotchkiss, Emily Kendall,  and Nathan Musoke for useful discussions and comments.  JCN acknowledges funding by a Julius von Haast Fellowship Award provided by the New Zealand Ministry of Business, Innovation and Employment and administered by the Royal Society of New Zealand. RE acknowledges support from the Marsden Fund of the Royal Society of New Zealand.

\providecommand{\href}[2]{#2}\begingroup\raggedright\endgroup

\end{document}